\documentclass{aa}

\usepackage{graphicx}
\usepackage{txfonts}
\bibpunct{(}{)}{;}{a}{}{,} 
\begin{document}

\title{K-shell photoabsorption and photoionization of trace elements}

\subtitle{II. Isoelectronic sequences with electron number $12\leq N\leq 18$}

\author{C. Mendoza\inst{1}\thanks{Also Emeritus Research Fellow, IVIC, Caracas, Venezuela.}
  \and
        M.~A. Bautista\inst{1}
  \and
         P. Palmeri\inst{2}
  \and
        P. Quinet\inst{2,3}
  \and
        M.~C. Witthoeft\inst{4,5}
  \and
        T.~R. Kallman\inst{5}
}

\institute{Department of Physics, Western Michigan University, 1903 W Michigan Ave., Kalamazoo, MI 49008, USA \\
               \email{claudio.mendozaguardia@wmich.edu, manuel.bautista@wmich.edu}
   \and
          Physique Atomique et Astrophysique, Universit\'e de Mons - UMONS, 20 place du Parc, 7000, Mons, Belgium \\
              \email{patrick.palmeri@umons.ac.be, pascal.quinet@umons.ac.be}
   \and
          IPNAS, Universit\'e de Li\`ege, Campus du Sart Tilman, B\^at. B15, 4000, Li\`ege, Belgium
   \and
          ADNET Systems, Inc., Bethesda, MD 20817, USA
   \and
          NASA Goddard Space Flight Center, Greenbelt, MD, 20771, USA \\
              \email{michael.c.witthoeft@nasa.gov, timothy.r.kallman@nasa.gov}
             }

\date{Received , ; accepted , }


\abstract
{We are concerned with improving the diagnostic potential of the K lines and edges of elements with low cosmic abundances, namely F, Na, P, Cl, K, Sc, Ti, V, Cr, Mn, Co, Cu, and Zn, that are observed in the X-ray spectra of supernova remnants, galaxy clusters and accreting black holes and neutron stars.}
{Since accurate photoabsorption and photoionization cross sections are needed in their spectral models, they have been computed for isoelectronic sequences with electron number $12\leq N\leq 18$ using a multi-channel method.}
{Target representations are obtained with the atomic structure code {\sc autostructure}, and ground-state cross sections are computed with the Breit--Pauli $R$-matrix method ({\sc bprm}) in intermediate coupling, including damping (radiative and Auger) effects.}
{Following the findings in our earlier work on sequences with $2\leq N\leq 11$, the contributions from channels associated with the 2s-hole $[2{\rm s}]\mu$ target configurations and those containing 3d orbitals are studied in the Mg ($N=12$) and Ar ($N=18$) isoelectronic sequences. Cross sections for the latter ions are also calculated in the isolated-resonance approximation as implemented in {\sc autostructure} and compared with {\sc bprm} to test their accuracy.}
{It is confirmed that the collisional channels associated with the $[2{\rm s}]\mu$ target configurations must be taken into account owing to significant increases in the monotonic background cross section between the L and K edges. Target configurations with 3d orbitals give rise to fairly conspicuous unresolved transition arrays in the L-edge region, but to a much lesser extent in the K-edge which is our main concern; therefore, they have been neglected throughout owing to their computationally intractable channel inventory, thus allowing the computation of cross sections for all the ions with $12\leq N\leq 18$ in intermediate coupling with {\sc bprm}. We find that the isolated-resonance approximations performs satisfactorily and will be our best choice to tackle the systems with ground configuration ${\rm 3p^63d}^m$ ($3\leq m\leq 8$) in isoelectronic sequences with $N > 20$.}


\titlerunning{K-shell photoabsorption of trace elements}
\authorrunning{C. Mendoza et al.}

\maketitle


\section{Introduction}

K lines and edges from elements with low cosmic abundance (trace elements) have been observed with the {\em Chandra}, {\em XMM-Newton} and {\em Suzaku} space telescopes in the X-ray spectra of supernova remnants, galaxy clusters, and accreting black holes and neutron stars. They have been used to constrain the ionization, metallicity, abundances, black-hole progenitor mass, and, in the case of warm absorbers, the flow properties \citep[see, for instance,][]{hwa00, mil06, bad08, kal09, tam09, ued09, nob10, par13}. Their diagnostic potential was set to become exploited with the launching last February 2016 of the {\em Hitomi} spatial telescope (previously referred to as {\em Astro-H}), an instrument of remarkable spectroscopic capabilities. In spite of its becoming inoperative soon after, there is currently a definite intention to replace it by 2020; therefore, the continuity of support projects, such as the present one to compute the atomic data required in specialised spectral modeling, is still arguable.

We have been involved for some time in the computation of atomic data to enable the reliable modeling of the trace-element K lines. \citet{pal12}, to be referred to hereafter as PQM12, computed with {\sc hfr} \citep[a Hartree--Fock code with relativistic corrections by][]{cow81} level energies, radiative and Auger widths, and fluorescence yields for K-vacancy levels of the complete isonuclear sequences of F, Na, P, Cl, K, Sc, Ti, V, Cr, Mn, Co, Cu, and Zn. Using the atomic target models developed therein, \citet{pal16} (Paper~1 hereafter) calculated photoabsorption and photoionization cross sections of the ground state of ionic systems with electron number $N\leq 11$. In this report, we extend this work to the third-row isoelectronic sequences  ($12\leq N\leq 18$) which, owing to the atomic complexity brought about by the open 3p subshell, implies a different computational strategy.

Cross sections for the fine-structure ground levels of the second-row ions ($N\leq 10$) were computed in Paper~1 in the close-coupling framework with the Breit--Pauli $R$-matrix method \citep[{\sc bprm},][]{ber95}, which enabled the rendering of the resonance structures of the open L shell and K ionization edge. Radiation and spectator-Auger damping that causes K-edge smearing was taken into account \citep{pal02}. The Na-like ions ($N=11$), in particular \ion{P}{v} and \ion{Zn}{xx}, were then used as test cases to extend this approach to the more complex members of the third row with $12\leq N\leq 18$. For instance, since the inclusion of target configurations with 3d orbitals dramatically increases the collisional channel inventory, particularly when the 3p subshell is half filled, $R$-matrix calculations would then have to be performed in $LS$ coupling rather than in intermediate coupling or, alternatively, be replaced by the uncoupled  isolated-resonance approximation \citep{abd12} of the atomic structure code {\sc autostructure} \citep{eis74, bad11}. We found in Paper~1 that 3d resonances were mainly conspicuous in the L edge rather than the K edge, the latter being our main concern and, therefore, they could be practically neglected in spite of the ever-present $n=3$ configuration-interaction (CI) effects.

These computational tradeoffs are examined in greater depth in this work, specifically for the simpler isoelectronic sequences with $N=12$ and $N=18$, in an attempt to maintain the intermediate coupling scheme for all the ionic systems of interest. A brief description of the numerical methods is given in Sect.~\ref{methods} and the results of the target-representation investigations and final calculations are discussed in Sect.~\ref{results}. Conclusions are drawn in Sect.~\ref{conc}, together with some guidelines for the future consideration of the fourth-row ions.


\section{Numerical methods}
\label{methods}

This project intends to compute photoabsorption and total and partial photoionization cross sections for ionic species of the isonuclear series P, Cl, K, Sc, Ti, V, Cr, Mn, Co, Cu, and Zn with electron numbers $N< Z-1$, where $Z$ is the atomic number identifying the sequence. Cross sections for isoelectronic sequences with $N\leq 11$ were reported in Paper~1, and here we discuss the details of those with $12\leq N\leq 18$. As previously mentioned in Paper~1, species with $Z-1\leq N\leq Z$ will be treated elsewhere.

The close-coupling framework of scattering theory is the main quantum-mechanical backbone, where the wave function of the total $N$-electron system is expanded in terms of the eigenfunctions of an ($N-1$)-electron target. Calculations are performed in intermediate coupling with the relativistic (Breit--Pauli) {\sc bprm} method, where configuration space is partitioned by a sphere of radius $r=a$ into two regions: an inner region ($r\leq a$) treated with the {\sc rmatrx1} package that takes into account exchange and correlation effects between the target and the active electron; and an asymptotic region ($r> a$) where these effects may be neglected and the active electron is subject to long-range multipole potentials. This external region is solved with the {\sc stgbf0damp7} code that also includes radiation and spectator-Auger damping by means of a model potential \citep{rob95, gor96, gor00}. For comparison purposes, particularly for the large target representations, the $R$-matrix package is also run in the more concise $LS$ coupling scheme.

Adopting the target representations listed in Table~9 of PQM12 -- namely configuration expansions, level energies, and Auger widths -- orbitals are generated in a Thomas--Fermi--Dirac statistical potential with the {\sc autostructure} atomic structure code \citep{eis74, bad11}. Cross sections in the isolated-resonance approximation are also obtained with this code. We note that, due to the complexity of third-row ions, the target representations of PQM12 do not list the levels that arise from a large number of $n=3$ correlation configurations that were taken into account to obtain reasonably accurate level energies and a level order similar to the spectroscopic.

\section{Results}
\label{results}

Cross sections are calculated right from threshold up to the monotonic decreasing tail beyond the K edge. Although our main interest is the K-edge structure, an effort is also made to resolve the resonances associated with the $n=3$ valence shell and the L edge. In the resonance regions, a fine mesh step of $E/z^2=0.0001$~Ryd is used where $z=Z-N+1$.

Following the findings of Paper~1, we analyse some of the problems discussed therein, in particular the contributions from $[2s]\mu$ target configurations and those from missing $n=3$ configurations (mainly those with 3d orbitals) in both the K- and L-edge regions using ions of the Mg ($N=12$) and Ar ($N=18$) isoelectronic sequences.

\subsection{$[2s]\mu$ target configurations}
\label{2s}

In Paper~1 we find that, in the case of the Na-like sequence ($N=11$), the corresponding Ne-like target models in PQM12 comprised the following configurations: ${\rm 2p}^6$, $[{\rm 2p]3s}$, $[{\rm 2p]3p}$, $[{\rm 1s]3s}$, and $[{\rm 1s]3p}$, but neglected $[{\rm 2s]3s}$ and $[{\rm 2s]3p}$. If the latter two are not taken into account in the close-coupling expansion, some  resonances (mostly narrow) in the L-edge region would be missing, but more importantly, the background cross section between the L and K edges is significantly underestimated. This effect is further examined here in Mg-like \ion{P}{iv} and Ar-like \ion{Sc}{iv} by using the following target representations:
\begin{description}
  \item[{\bf Target A} --] As listed in Table~9 of PQM12, the Na-like \ion{P}{v} target is represented with the configurations ${\rm 2p^63s}$, ${\rm 2p^63p}$, $[{\rm 2p}]{\rm 3s^2}$, $[{\rm 2p}]{\rm 3s3p}$, $[{\rm 1s}]{\rm 3s^2}$ and $[{\rm 1s}]{\rm 3s3p}$;

  \item[{\bf Target B} --] The \ion{P}{v} target includes the configurations of Target~A plus $[{\rm 2s}]{\rm 3s^2}$ and $[{\rm 2s}]{\rm 3s3p}$;

  \item[{\bf Target C} --] As listed in Table~9 of PQM12, the Cl-like \ion{Sc}{v} target is represented with the configurations ${\rm 3p}^5$, $[{\rm 3s]3p^6}$, $[{\rm 2p]3p^6}$ and $[{\rm 1s]3p^6}$;

  \item[{\bf Target D} --] \ion{Sc}{v} target includes the configurations of Target~C plus $[{\rm 2s]3p^6}$.
\end{description}

\begin{figure}[t]
  \centering
  \includegraphics[width=8.0cm]{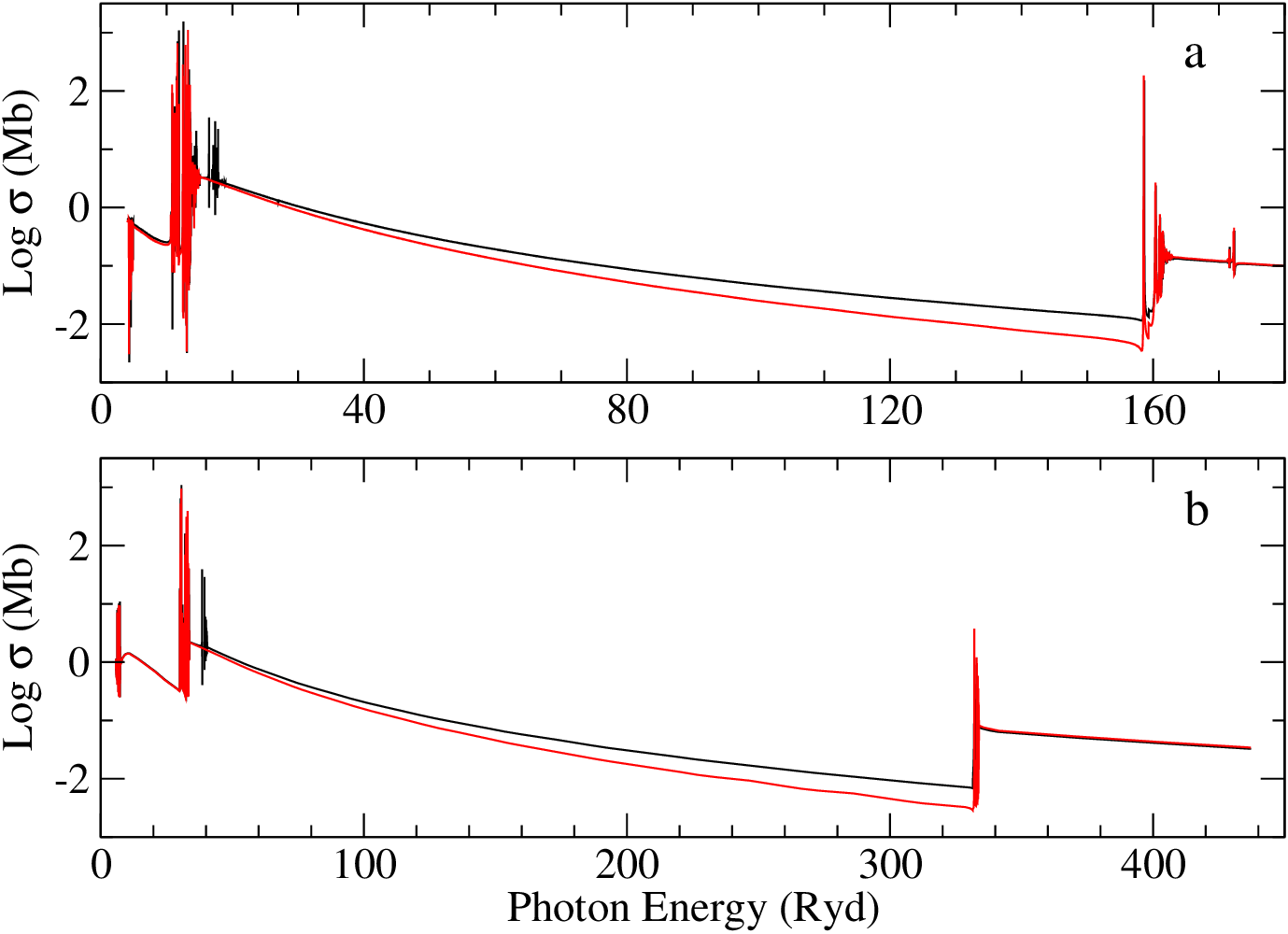}
  \caption{Photoabsorption cross sections computed with {\sc bprm} in intermediate coupling of (a) Mg-like \ion{P}{iv} (Target~A, red curve; Target~B, black curve) and (b) Ar-like \ion{Sc}{iv} (Target~C, red curve; Target~D, black curve). Underestimated background cross sections between the L and K edges are obtained when the $[{\rm 2s}]\mu$ target configurations are excluded (red curves).} \label{bg-cs}
\end{figure}

In Fig.~\ref{bg-cs} the photoabsorption cross sections of \ion{P}{iv} and \ion{Sc}{iv} ions are plotted in both the L- and K-edge regions, where the underestimated background cross sections caused by the exclusion of channels associated to the $[{\rm 2s}]\mu$ target configurations can be seen. Although the K edge is hardly altered, when these configurations are taken into account, additional unresolved transition arrays (UTAs) appear ($\sim 16.5$~Ryd for \ion{P}{iv} and $\sim 38.5$~Ryd in \ion{Sc}{iv}) in the L-edge region.


\begin{figure}[h]
\vspace{7mm}
  \centering
  \includegraphics[width=8.0cm]{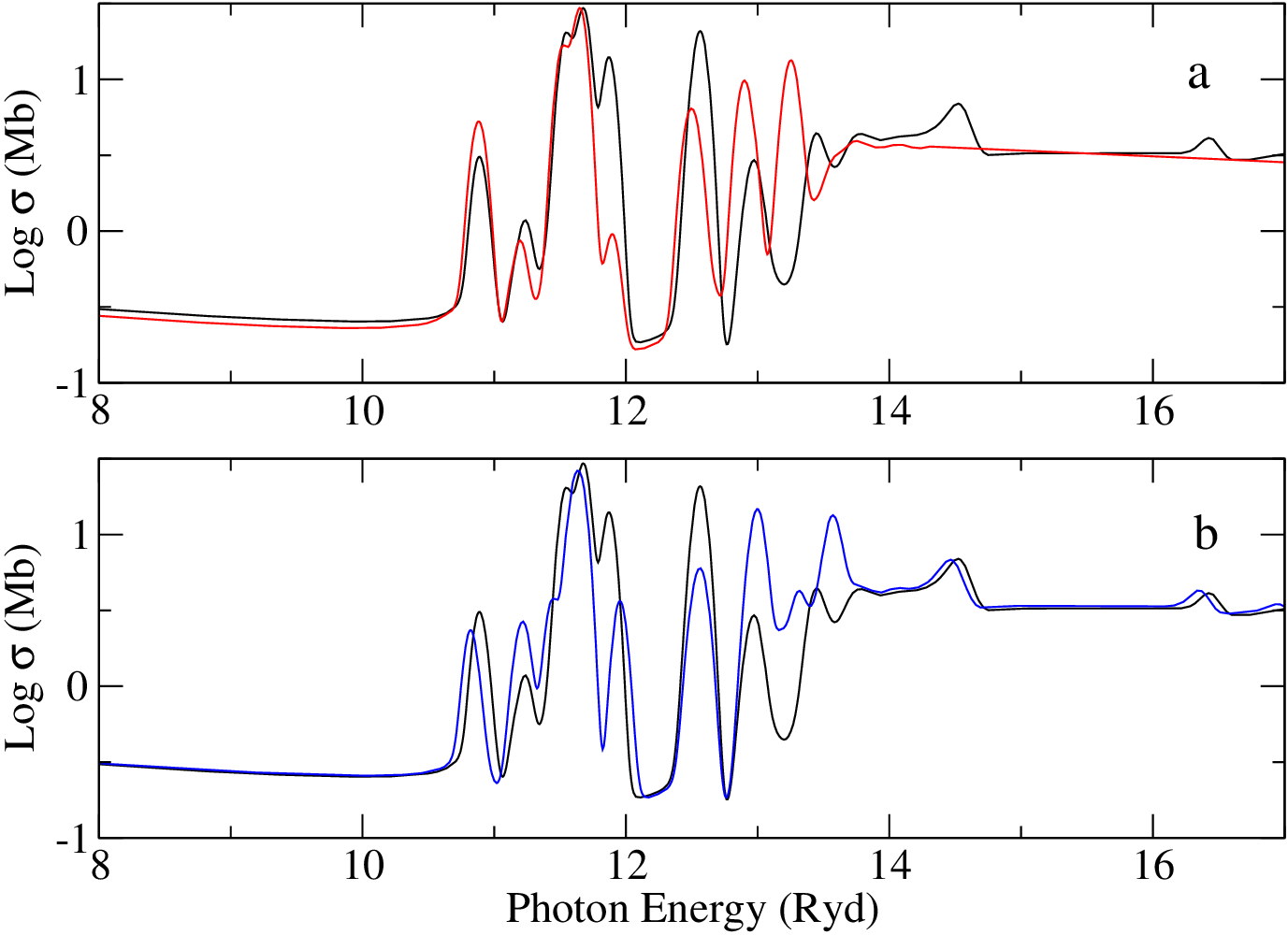}
  \caption{Photoabsorption cross section  of the ground state of Mg-like \ion{P}{iv} computed with {\sc bprm} in intermediate coupling in the L-edge region. (a) Red curve, Target~A; black curve, Target~B. (b) Black curve, Target~B; blue curve,  Target~E. The cross sections have been convolved with a Gaussian of width $\Delta E/E=0.01$.} \label{L-edge}
\end{figure}

In the case of \ion{P}{iv}, the L edge is amplified in Fig.~\ref{L-edge}a where, to estimate the variations of the UTA morphology, the cross sections have been convolved with a Gaussian of width $\Delta E/E=0.01$. We see that, apart from the extra features emerging when using Target~B in the 14--17~Ryd energy interval, the resonance structure is also somewhat perturbed around the edge head ($\sim 13$~Ryd).


\subsection{Missing $n=3$ configurations}
\label{n3}

An inherent problem in the computation of radiative properties for third-row ions ($11\leq N\leq 18$) with an incomplete 3p subshell is the large number of $n=3$ closely coupled states to be considered from configurations of the type ${\rm 3s}^x{\rm 3p}^y{\rm 3d}^z$. If electron promotions from the K and L shells are additionally involved, such as in this work, then the channel inventory in the close-coupling expansion rapidly becomes computationally intractable; therefore, compromises are unavoidable in spite of the ensuing CI shortcomings. For instance, in the Na-like targets ($N=11$) of PQM12, configurations of the type $[{\rm 1s}]{\rm 3p}^2$ and $[{\rm 2p}]{\rm 3p}^2$ were neither listed, nor those containing 3d orbitals for all the systems with $N\leq 18$.

To study the impact of these exclusions on the edge-region cross sections, three targets are considered:
\begin{description}
  \item[{\bf Target E} --] Na-like \ion{P}{v} target is represented with configurations of Target~B plus $[{\rm 2p}]{\rm 3p^2}$, $[{\rm 2s}]{\rm 3p^2}$ and $[{\rm 1s}]{\rm 3p^2}$;

  \item[{\bf Target F} --] \ion{P}{v} target includes the configurations of Target~E plus ${\rm 2p^63d}$, $[{\rm 2p}]{\rm 3s3d}$, $[{\rm 2p}]{\rm 3p3d}$, $[{\rm 2p}]{\rm 3d^2}$, $[{\rm 2s}]{\rm 3s3d}$, $[{\rm 2s}]{\rm 3p3d}$, $[{\rm 2s}]{\rm 3d^2}$, $[{\rm 1s}]{\rm 3s3d}$, $[{\rm 1s}]{\rm 3p3d}$ and $[{\rm 1s}]{\rm 3d^2}$;

  \item[{\bf Target G} --] Cl-like \ion{Sc}{v} target is represented with the configurations of Target~D plus ${\rm 3s^23p^43d}$, ${\rm 3s3p^53d}$, $[{\rm 2p]3s^23p^53d}$, $[{\rm 2p]3s3p^63d}$, $[{\rm 2s]3s^23p^53d}$, $[{\rm 2s]3s3p^63d}$, $[{\rm 1s]3s^23p^53d}$ and $[{\rm 2s]3s3p^63d}$.
  \end{description}
Target~B comprises 39 fine-structure levels, an inventory that is almost doubled in Target~E to 76 levels while Target~F gives rise to 305. This sharp level increase, when only single-electron promotions are considered, supports our previous comments on computationally unmanageable atomic models, even for the relatively simple Na- and Cl-like targets. A similar situation is found for the latter, where Target~D encompasses 72 levels in comparison with 208 in Target~G.

\begin{figure}[h!]
\vspace{7mm}
  \centering
  \includegraphics[width=8.0cm]{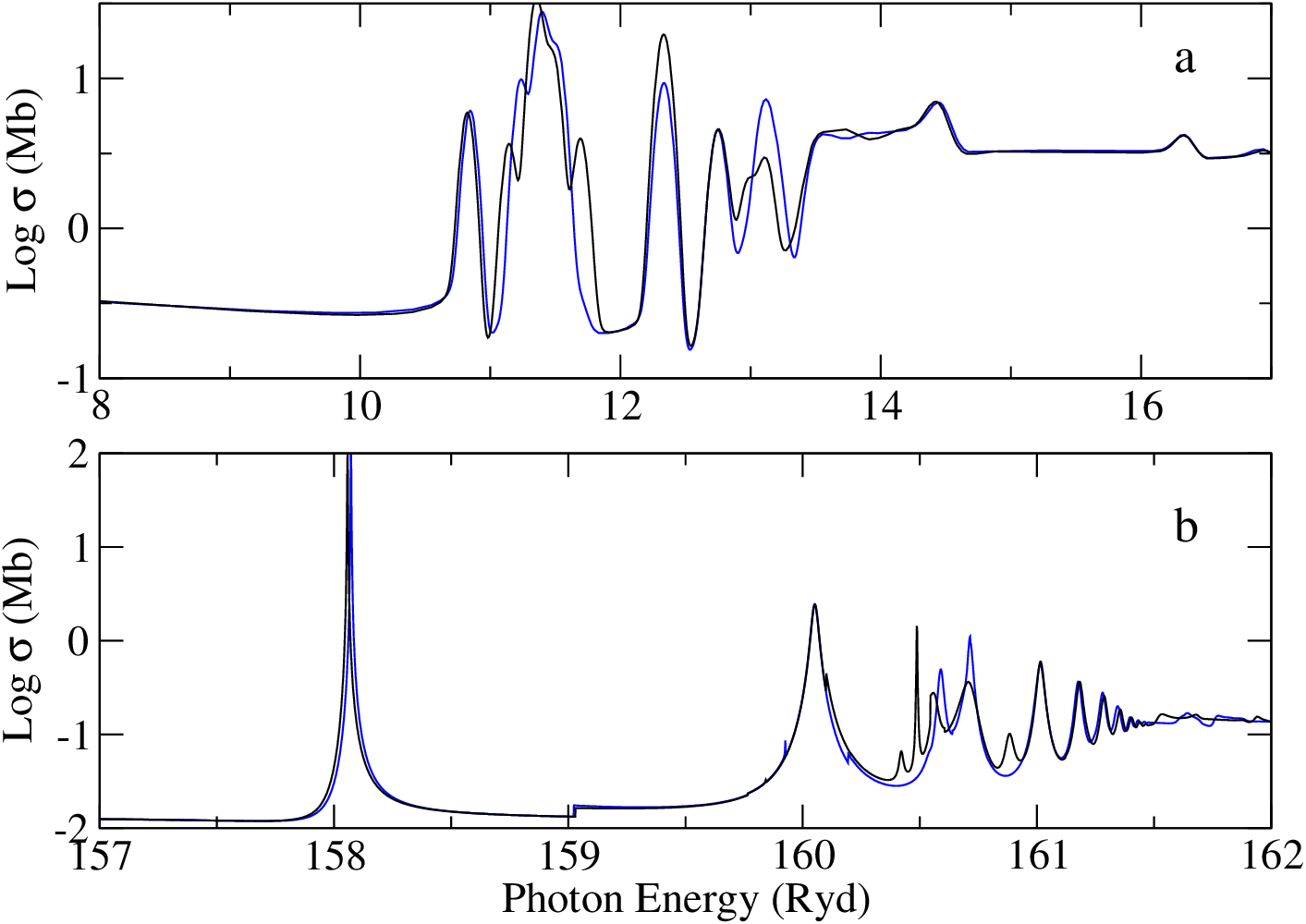}
  \caption{Photoabsorption cross section of the ground state of Mg-like \ion{P}{iv} computed with {\sc bprm} in $LS$ coupling in the (a) L-edge (convolved with a Gaussian of width $\Delta E/E=0.01$) and (b) K-edge regions. Blue curve, Target~E. Black curve, Target~F. \label{kl1}}
\end{figure}

\begin{figure}[t!]
\vspace{7mm}
  \centering
  \includegraphics[width=8.0cm]{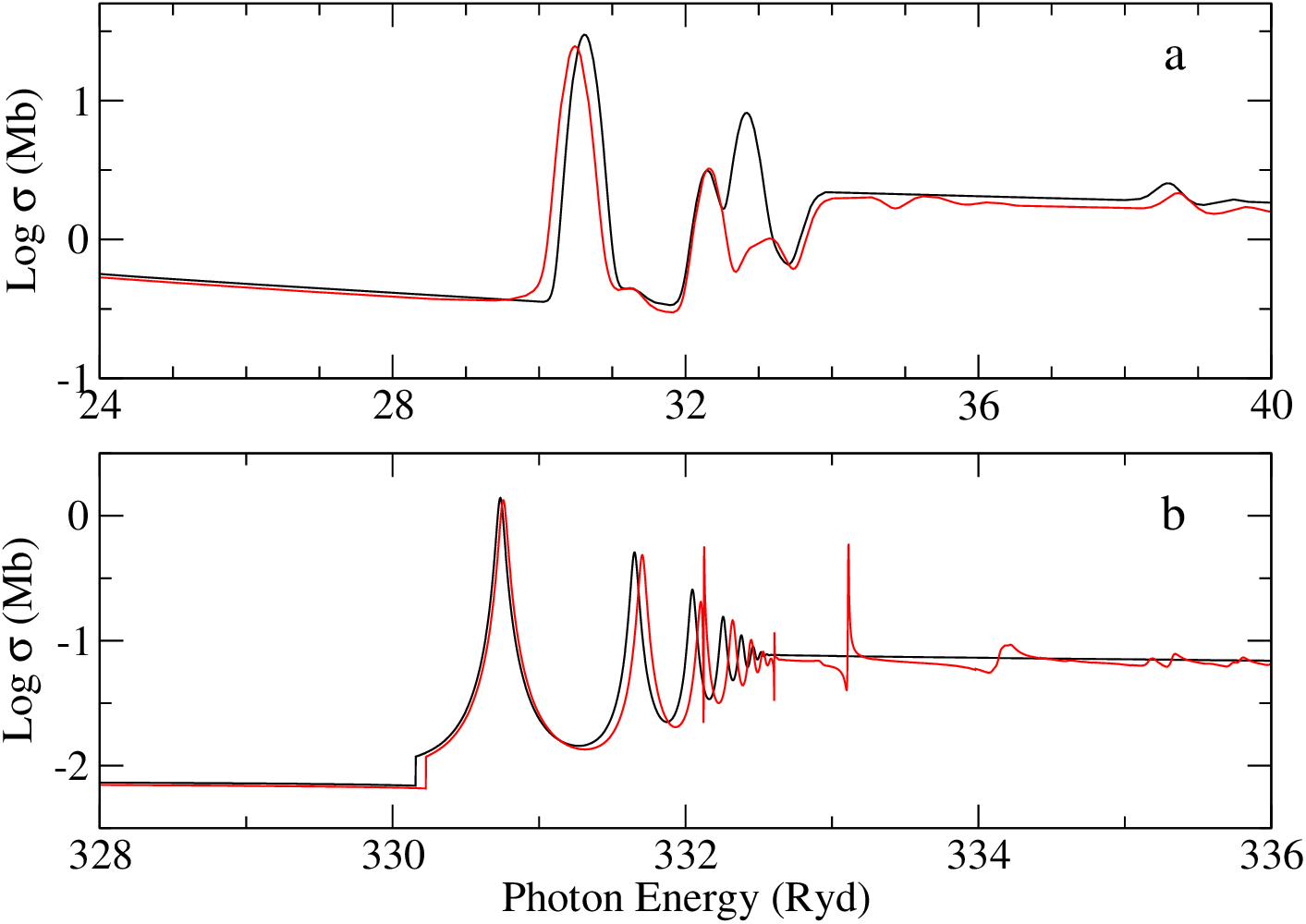}
  \caption{Photoabsorption cross sections of the ground state of Ar-like \ion{Sc}{iv} computed with {\sc bprm} in $LS$ coupling in the (a) L-edge (convolved with a Gaussian of width $\Delta E/E=0.01$) and (b) K-edge regions. Black curve, Target~D. Red curve, Target~G. \label{kl2}}
\end{figure}

Photoabsorption cross sections for Mg-like \ion{P}{iv} in the K-edge region computed with Target~B and Target~E are very similar but, as shown in Fig.~\ref{L-edge}b, the UTAs in the edge head at 13--14~Ryd show different morphologies. These behavioral patterns are maintained in comparisons between the cross sections computed with Target~E and Target~F (Fig.~\ref{kl1}) and, in the case of Ar-like \ion{Sc}{iv}, between Target~D and Target~G (Fig.~\ref{kl2}). Owing to the large number of collisional channels, cross sections in Figs.~\ref{kl1}--\ref{kl2} have been computed in $LS$ coupling; thus, a comparison between the black curves in Fig.~\ref{L-edge}b and Fig.~\ref{kl1}a gives a measure of the small but distinguishable resonance-structure differences resulting from the $LS$- and intermediate-coupling frameworks.


\begin{figure}[h!]
\vspace{7mm}
  \centering
  \includegraphics[width=8.4cm]{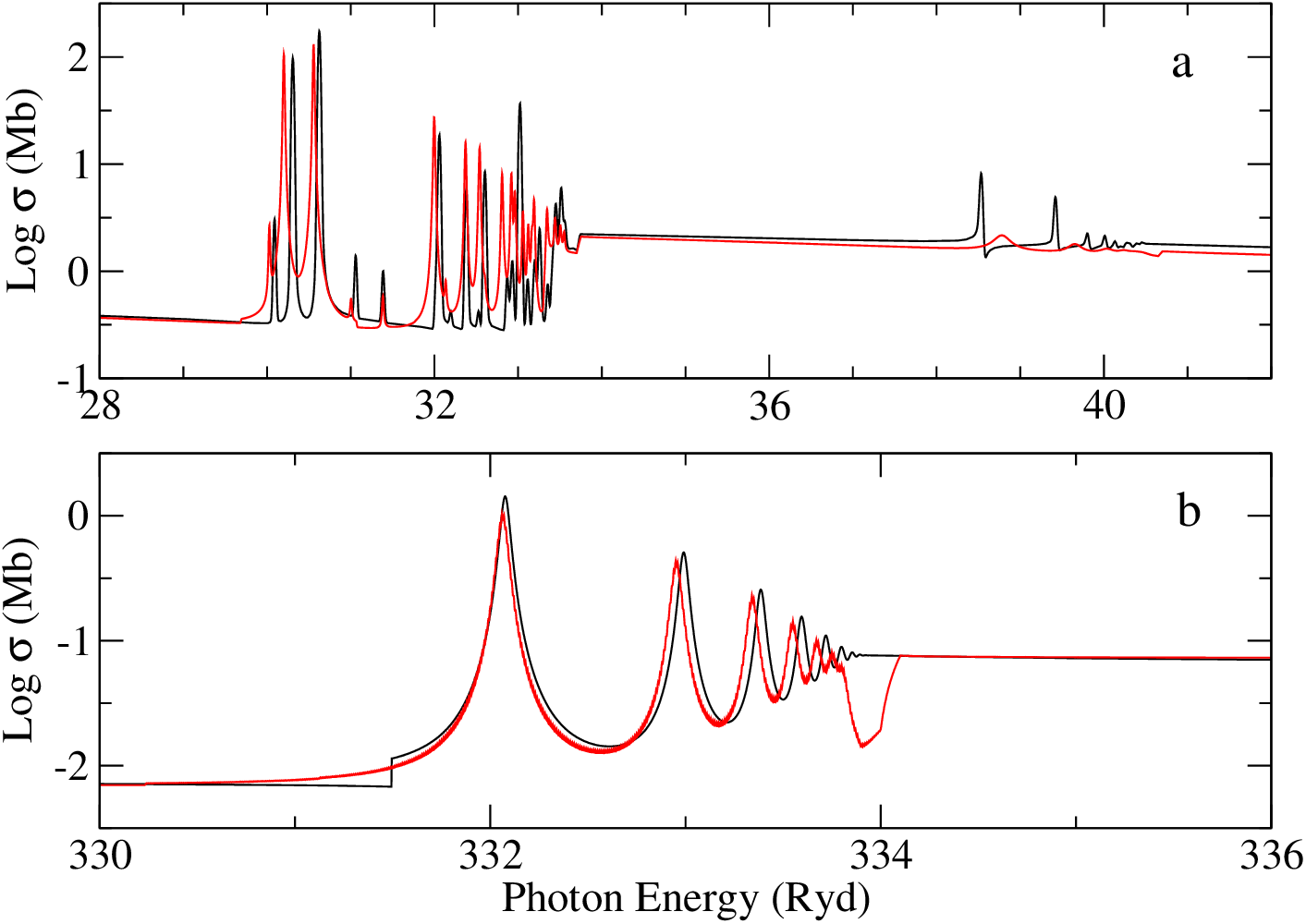}
  \caption{Intermediate-coupling photoabsorption cross sections of the ground state of Ar-like \ion{Sc}{iv} in the (a) L-edge (convolved with a Gaussian of width $\Delta E/E=0.001$) and (b) K-edge regions. Black curve, computed with {\sc bprm} with Target~D. Red curve, computed with Target~D in the isolated resonance approximation with {\sc autostructure}. \label{autos}}
\end{figure}


\subsection{Isolated-resonance approximation}

A comparison is depicted in Fig.~\ref{autos} between the photoabsorption cross sections of \ion{Sc}{iv} in the K- and L-edge regions computed in intermediate coupling with {\sc bprm} and, in the isolated resonance approximation, with {\sc autostructure}, where it may be seen that the latter performs surprisingly well in spite of the neglect of channel coupling and the adoption of symmetric Lorentzian resonance profiles. In this plot, a small energy shift was performed to match the resonance series, which is due to differences in the {\sc bprm} and {\sc autostructure} ionization potentials. Moreover, a comparison of the  $[{\rm 1s}]n{\rm p}\,^1{\rm P^o}$ K resonance position in Fig.~\ref{kl2}b and Fig.~\ref{autos}b reveals a 1.5~Ryd discrepancy that is caused by the relativistic contribution since the former was obtained in $LS$ coupling. This discrepancy can be reduced by running {\sc bprm} in $LS$ coupling but including the relativistic corrections that do not lead to fine structure; i.e. the mass-correction and Darwin one-body terms.

\subsection{Sequences with $13\leq N\leq 17$}

Taking into account the findings that emerged from the different target representations considered for the Mg- and Ar-like isoelectronic sequences in Sects.~\ref{2s}--\ref{n3}, we proceeded to compute with {\sc bprm} cross sections for the other more computationally involved sequences, namely those with electron number $13\leq N\leq 17$. We adopted the target models of PQM12, but additionally included the $[{\rm 2s}]\mu$ configurations that, as shown in Fig.~\ref{bg-cs}, affect the background photoabsorption cross sections between the L and K edges. Configurations including 3d orbitals were thus excluded leading to underrepresented L edges (see Figs.~\ref{kl1}--\ref{kl2}).

\begin{figure*}[t!]
  \centering
  \includegraphics[width=16cm]{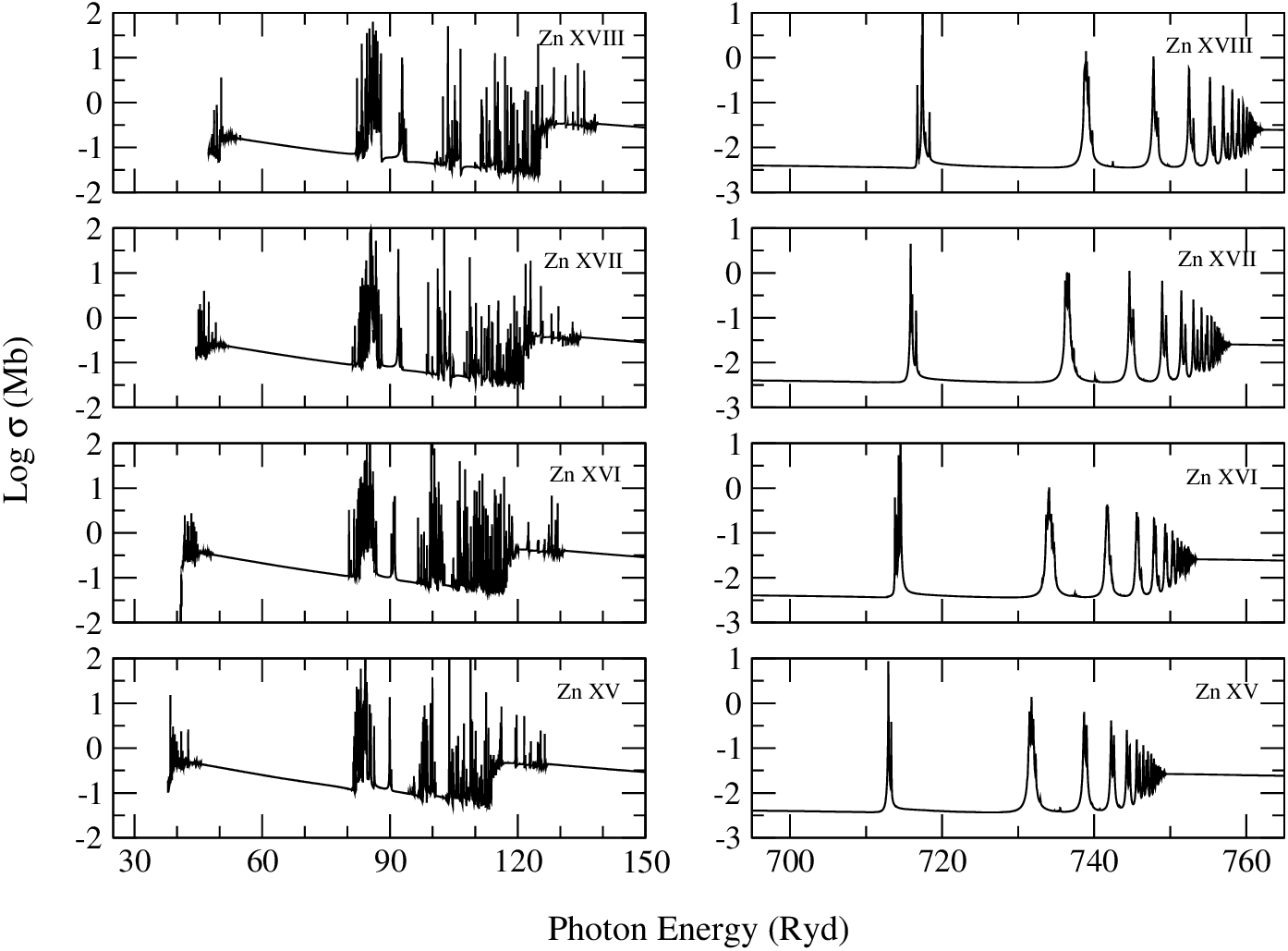}
  \caption{{\sc bprm} intermediate-coupling photoabsorption cross sections of the ground states of  \ion{Zn}{xviii} (Al-like),  \ion{Zn}{xvii} (Si-like),  \ion{Zn}{xvi} (P-like) and  \ion{Zn}{xv} (S-like) in the L- (left column) and K-edge (right column) regions. \label{mosaic}}
\end{figure*}

As an example, photoabsorption cross sections in the K- and L-edge regions for Zn ions with electron number $14\leq N\leq 17$ are plotted in Fig.~\ref{mosaic}, showing contrasting edge resonance structures. The K edge is dominated by well-defined resonance series (K lines) whose broad and practically constant widths are regulated by damping (radiative and spectator-electron Auger); they thus lead to smeared edges in the photon energy band 745--765~Ryd. The sharp L edges in the energy band 110--130~Ryd are populated by a very large number of narrow resonances where the $2\rightarrow 3$ and $2\rightarrow 4$ UTAs at $\sim 85$~Ryd and $\sim 100$~Ryd, respectively, can be distinguished. Less pronounced UTAs between 38 and 55 Ryd due to valence resonances may also be appreciated.

From the astronomical diagnostic point of view, Fig.~\ref{mosaic} appears to indicate that the K$\alpha$, K$\beta$, and K edge will give rise to resolvable spectral signatures with information about the plasma ionization state, while L edge dominated by several UTAs will be messy and thus less astrophysically useful; in fact, in such spectral ranges K lines from the lower-$Z$ species are the features to single out.

\section{Conclusions}
\label{conc}

\begin{enumerate}

  \item Photoabsorption and photoionization (total and partial) cross sections for the ions of P, Cl, K, Sc, Ti, V, Cr, Mn, Co, Cu, and Zn with electron numbers $12\leq N\leq 18$ have been computed with the Breit--Pauli $R$-matrix method ({\sc bprm}) in intermediate coupling. Both the L and K resonance structures are well resolved with fine energy meshes. Damping effects have been shown to be relevant in the K-edge resonance structures.

  \item Regarding the importance of 2s promotion discussed in Paper~1, it is reaffirmed that, for third-row ions, this must be taken into account with configurations of the type $[{\rm 2s}]\mu$ otherwise the background cross sections between the K and L edges are underestimated.

  \item We have excluded target configurations with 3d orbitals, thus enabling the computations of cross sections in intermediate coupling with {\sc bprm} that adopt the ion models of PQM12 for isoelectronic sequences with $13\leq N\leq 17$. As shown in Sect.~\ref{n3}, the L edge will be somewhat affected, but a more realistic representation of its intricate resonance series would require more sophisticated target models and a computational effort beyond the scope of this project that is mainly focused on the K-edge structures. For tractable target-model management, we have only considered single excitations within the $n=3$ complex that certainly lead to unconverged CI expansions, in particular in the valence electron shell. In Paper~1, the possibility of rendering the K and L edges in separate calculations was mentioned, but this scheme was dropped since it would have made the determination of partial photoionization cross section, which are required in the atomic database of the {\sc xstar} spectral modeling code \citep{bau01, kal01}, very cumbersome.

  \item For ions with $N=18$, namely \ion{Sc}{iv}, we have made comparisons of the present {\sc bprm} photoabsorption cross sections with those computed in the isolated-resonance approximation with the {\sc autostructure} code. This is an important step since cross sections for most of the lowly ionized species with electron number $N>20$ and ground configurations ${\rm 3p^63d}^m$ with $3\leq m\leq 8$ will not be calculated with {\sc bprm} owing to target size. In this respect, it has been shown that, in spite of the shortcomings of the isolated-resonance approximation, i.e. neglected channel-coupling and symmetric resonance profiles, its performance and general accuracy is satisfactory.

  \item When cross sections are computed with {\sc bprm} in $LS$ coupling, it is emphasized that the relativistic one-body, non-fine-structure corrections must be taken into account otherwise poor energy resonance positions, in particular those associated with the K edge, will be obtained.

  \item Data access will be managed through the Centre de Donn\'ees astronomiques de Strasbourg (CDS\footnote{http://cdsweb.u-strasbg.fr/}) taking into consideration previous long-term arrangements with this data center, which have led to sustainable and efficient data services, e.g. TOPbase\footnote{http://cdsweb.u-strasbg.fr/topbase/topbase.html} \citep{cun93}. The complete data sets will be uploaded with the upcoming Paper~3 of the present series, and the corresponding links will be provided through the ADS.
\end{enumerate}

\begin{acknowledgements}
This project is sponsored by the NASA grant 12-APRA12-0070 through the Astrophysics Research and Analysis Program. Pascal Quinet and Patrick Palmeri are Research Director and Research Associate, respectively, of the Belgian Fund for Scientific Research F.R.S.-FNRS.
\end{acknowledgements}

\bibliographystyle{aa}


\end{document}